\documentstyle{article}
\begin{document}
{\large

\begin{center}
\vspace{1cm}
{\large \bf
Overlap integaral fot quantum skyrmions
}

\vspace{0.5cm}
{R. A. Istomin and A. S. Moskvin}

\vspace{0.3cm}
{\small
Department of Theoretical Physics, Ural State University,
620083, Lenin Ave. 51, Ekaterinburg , Russia.
}
\vspace{0.5cm}

\end{center}

\begin{abstract}
Making use of the method of spin coherent states we have obtained an analytical
form for overlap integral for quantum skyrmions.
\end{abstract}

PACS numbers 75.10.Jm, 03.65.Sq

Skyrmions are general static solutions of  2D Heisenberg ferromagnet, obtained
by Belavin and Polyakov \cite{BP} from classical nonlinear sigma model. A
renewed interest to these unconventional spin textures is stimulated by
high-$T_c$ problem in doped quasi-2D-cuprates and  quantum Hall effect.

The spin distribution within classical skyrmion of topological charge $q=1$ is
given as follows
\begin{equation}
S_{x}=\frac{2r\lambda }{r^{2}+\lambda ^{2}}\cos \varphi ,\qquad
S_{y}=\frac{2r\lambda }{r^{2}+\lambda ^{2}}\sin \varphi ,\qquad
S_{z}=\frac{r^{2}-\lambda ^{2}}{r^{2}+\lambda ^{2}}.
\end{equation}
In terms of the stereographic variables the skyrmion with radius  $\lambda $
and phase  $\varphi _{0}$  centered at a point $z_0$ is identified with spin
distribution  $w(z)=\frac{\Lambda }{z-z_0}$, where $z=x+iy=re^{i \varphi }$  is
a point in the complex plane, $\Lambda =\lambda e^{i\theta}$, and characterised
by three modes: translational $z_0$-mode,  "rotational" $\theta$-mode and
"dilatational" $\lambda$- mode. Each of them relates to  certain symmetry of
the classical skyrmion configuration. For instance, $\theta$- mode corresponds
to combination of rotational symmetry and internal $U(1)$ transformation.

The simplest wave function of the spin system, which corresponds to classical
skyrmion,  is a product of spin coherent states \cite{Perelomov}. In case of
spin $s=\frac{1}{2}$
\begin{equation}
\Psi _{sk}( 0) =\prod\limits_{i}\lbrack \cos \frac{\theta
_{i}}{2}e^{i\frac{\varphi _{i}}{2}}\mid \uparrow \rangle +\sin \frac{\theta
_{i}}{2}e^{-i\frac{\varphi _{i}}{2}}\mid \downarrow \rangle \rbrack ,
\end{equation}
where $\theta _{i}=\arccos \frac{r_{i}^{2}-\lambda ^{2}}{r_{i}^{2}+\lambda
^{2}}$. Coherent state implies a maximal equivalence to classical state with
minimal uncertainty of spin components. In this connection we should note that
such a state was used in paper \cite{Schliemann} by Schliemann and Mertens
where an expression for the square variance of the Heisenberg Hamiltonian was
obtained.

Classical skyrmions with different phases and radia have equal energy.
Nevertheless, stationary state of quantum skyrmion of topological charge $q=1$
is not a superposition of states with different phase and radius \cite{Stern},
but has a certain distinct value of $\lambda$.

In this paper  we consider some peculiarities of quasiparticle
behaviour for the quantum skyrmion. First of all it implies the calculation of
the overlap and "resonance" integrals:
$$
S_{12}=\langle \Psi _{sk}({\vec R}_{1})|\Psi _{sk}({\vec R}_{2}) \rangle,
\qquad H_{12}=\langle  \Psi _{sk}({\vec R}_{1})|\hat H| \Psi _{sk}({\vec
R}_{2}) \rangle .
$$
As a helpful illustration to the calculation of the overlap integral for
quantum topological defects  it is worth to present some known results
concerning the  overlap of vortices in 2D superconducting condensate
\cite{Niu2}.  Phenomenologically such vortices were described as point
quasiparticles, moving under the action of the transverse Magnus force.
Coherent state of the vortex with the center in ${\vec R}_{0}$ was taken in
 the form \cite{Niu2}
\begin {equation}
| \Psi _{sk}({\vec R}_{0})\rangle =\frac{1}{\sqrt{2\pi l^{2}}}\exp \bigl[
-\frac{|{\vec r}-{\vec R}_{0}|^{2}}{4l^{2}}+\frac{i{\vec z}\cdot {\vec
R}_{0}\times {\vec r}}{2l^{2}}\bigr],
\end {equation}
where  $\rho _{0}$  is a density of  2D condensate, $l=(2\pi \rho _{0})^{-1/2}$
is an average distance between particles in the condensate, ${\vec z}$ a unit
vector normal to the plane of condensate.  Overlap integral for two coherent
states is then easily calculated as
\begin{equation}
S_{12}=\langle\Psi _{sk}(\vec{R_{1})}|\Psi _{sk}( \vec{R_{2}})
\rangle=\exp[-\frac{1}{4l^{2}}(R_{12}^{2}-4i\Delta _{12})],
\end{equation}
where $$ \Delta
_{12}=\frac{1}{2}\vec{z}\cdot[\vec{R_{1}}\times\vec{R_{2}}],\quad
  R_{12}=|\vec{R_{1}}-\vec{R_{2}}|.
$$
 It  contains Gaussian factor
reflecting localisation of the coherent state and also a specific phase factor
$ \Delta _{12}, $ being the area of a sector topological defect  covers while
moving in the plane. This factor originates from the phase factor in the
function (3) typical for the charged particle moving in the magnetic field or,
in general, for particle which experiences a Magnus force. The correct form of
such a Magnus force for the out-of-plane magnetic vortex with topological
charge $\frac{1}{2}$ ("half-skyrmion") was derived by Nikiforov and Sonin
\cite{NS}, the Magnus force acting on the classical skyrmion is simply twice
larger \cite{Volovik}.

It seems, the expression (4) reflects two common  features of the overlap
integral for topological defects in 2D system, namely Gaussian dependence on
$R_{12}$ and presence of the Berry phase. One should note that in recent paper
by Thang \cite{Tang} it is shown that wave function (3) does not provide
correct description of the transition to the infinite system, when the overlap
integral turns to zero. Nevertheless, the expressions (3) and (4) for the wave
function and overlap integral allow to elucidate many generic features of the
corresponding quantities for the topological defects.

To reveal some peculiarities of the quantum quasiparticle behaviour for skyrmions
we consider overlap integral for the simple quantum state like (2) of spin
system with skyrmion  $\frac{\lambda }{z-R_{1}}$ at the point  $R_{1}=\mid
R_{1}\mid e^{i\varphi _{1}}$   with the state $\frac{\lambda }{z-R_{2}}$, which
corresponds to skyrmion shifted to an arbitrary distance
$$
R_{1}^{2}=\mid R_{1}\mid ^{2}+\mid R_{2}\mid^{2}-2R_{1}R_{2}\cos ( \varphi
_{1}-\varphi _{2})
$$
into the point $R_{2}=\mid R_{2}\mid e^{i\varphi _{2}}$. Overlap of the single
spin coherent states characterised by  two different points on the complex
plane $\varsigma$   and  $\mu$, is \cite{Perelomov}
\begin{equation}
\langle\varsigma|\mu \rangle= \frac{( 1+\varsigma \overline{\mu }) ^{2S}}{(
1+\mid \varsigma \mid ^{2}) ^{S}( 1+\mid \mu \mid ^{2}) ^{S}}.
\end{equation}
Thus, overlap integral for the skyrmion states is given  by
$$
S_{12}^{\lambda \lambda}=\prod\limits_{i} \frac{( 1+\varsigma \overline{\mu })
^{2S}}{( 1+\mid \varsigma \mid ^{2}) ^{S}( 1+\mid \mu \mid ^{2}) ^{S}}
=\exp(S\sum\limits_{i}\ln \lbrack \frac{( 1+\varsigma \overline{\mu }) ^{2}}{(
1+\mid \varsigma \mid ^{2}) ( 1+\mid \mu \mid ^{2}) }\rbrack )=
$$
$$
\exp(S\sum\limits_{i}\lbrack 2\ln ( 1+\frac{\lambda
^{2}}{r_{i}^{2}+R_{1}\overline{R_{2}}-r_{i}R_{1}e^{-i\varphi
_{i}}-r_{i}\overline{R_{2}}e^{+i\varphi _{i}}}
$$
$$
-\ln ( 1+\frac{\lambda ^{2}}{r_{i}^{2}+R_{2}^{2}-2R_{2}r_{i}\cos \varphi _{i}})
-\ln ( 1+\frac{\lambda ^{2}}{r_{i}^{2}+R_{1}^{2}-2R_{1}r_{i}\cos \varphi _{i}})
\rbrack )
$$
\begin{equation}
= \exp(2I(R_{1}, R_{2})-I(R_{1},R_{1})-I(R_{2},R_{2}).
\end{equation}
It should be noted that  for  displacements $R\geq2\lambda $ on continuous
complex plane the  overlap integral turns to zero since in this case there
exist two such points for which initial and final spin states are orthogonal.
For example, in case of $R_{1}=0, R_{2}=R$, where R is real, $1+\frac{\lambda
^{2}}{z( z-R) }=0$, when $z=\frac{R}{2}\pm \sqrt{\frac{R^{2}}{4}-\lambda
^{2}}$. However,  overlap of the skyrmion on the lattice turns to zero only at
certain values of $R$, which form a certain discrete set of values. In
particular, for  displacement  along the direction of elementary vector of the
lattice overlap is identically zero at the points
$R=\frac{\lambda^{2}+k^{2}}{k}$ given   integer k.

  To consider the skyrmions of large radius we turn the sum in the exponent of Eq. (6) into integral.
The quantity $\frac{c}{\lambda}$ for  large skyrmions  is small and virtually
we need to keep terms of zero order in $\frac{c}{\lambda}$, with $c$ being
lattice constant. Spin density,  that is the number of spins in the unit cell
of the plane,  is simply $\frac{1}{c^{2}}$. Theory will be invariant with
respect to scale transformations, that is to variations of  $c$ and
simultaneous equivalent variation of $\lambda$ and $R$. Besides, we consider
size $L$ of the system to be much larger than skyrmion size and virtually keep
only terms of zero order in $\lambda/L$. To this end we may simply set
$L=\infty$. So, the result is \begin{equation}
I(R_{1},
R_{2})=\frac{1}{c^{2}}\int\limits_{0}^{\infty } A(r,R_{1},R_{2})\,r\,dr,
\end{equation}
where
\begin{equation}
A(r,R_{1},R_{2})=\int\limits_{0}^{2\pi }d\varphi\,\ln (
\lambda^{2}+r^{2}+R_{1}\overline{R_{2}}-rR_{1}e^{-i\varphi
}-r\overline{R_{2}}e^{+i\varphi}).
\end{equation}
 In order to perform the angular integration in $A(r,R_{1},R_{2})$, we introduce complex variable $z=e^{+i\varphi}$, so that
\begin {equation}
A(r,R_{1},R_{2})=\int dz \frac{ln(
\lambda^{2}+r^{2}+R_{1}\overline{R_{2}}-rR_{1}z^{-1}-r\overline{R_{2}}z)}{iz},
\end {equation}
where the integration is taken on the circle of unit radius. This integral
appears to be nonzero due to   the pole in $z=0$ and the nonanaliticity area
connected with the existence of the cut in the space of values of the complex
logarithm when its argument is negative. While crossing the axis of negative
values of the argument the phase jumps from $-\pi$ to $\pi$. In the plane of
$z$ nonanaliticity area is the curve given by the equation
$$
Im(\lambda^{2}+r^{2}+R_{1}\overline{R_{2}}-rR_{1}z^{-1}-r\overline{R_{2}}z)=0,
$$
$$
Re(\lambda^{2}+r^{2}+R_{1}\overline{R_{2}}-rR_{1}z^{-1}-r\overline{R_{2}}z)<0.
$$
When  $R>2\lambda$ there exist interval of the values   $r$, for which  the end
of the cut lies on the border of the unit circle resulting in purely imaginary
contribution into integral $ A(r,R_{1},R_{2})$. The range of  $r$, in which
this occurs is given by the interval $[r_{1},r_{2}]$, where
$$
 r_{1,2}^{2}=(\mid R_{1}\mid^{2}+\mid R_{1}\mid^{2})/2-\lambda^{2}\pm
(\mid R_{1}\mid^{2}-\mid
R_{1}\mid^{2})\sqrt{\frac{1}{4}-\frac{\lambda^{2}}{R^{2}}}.
$$
As a result of existence of the cut we have the following expression for the
integral $ A(r,R_{1},R_{2})$
$$
 A(r,R_{1},R_{2})=2\pi
 \Biggl \{
  \begin{array}{ll}
    ln(
    \frac{\lambda^{2}+r^{2}+R_{1}\overline{R_{2}}+
    \sqrt{(\lambda^{2}+r^{2}+R_{1}\overline{R_{2}})^{2}-4r^2R_{1}\overline{R_2}}}
    {2}
    ),
  &
    \mbox{if } r>r_{2} \mbox{ or } r<r_{1} \\
    ln(rR_{1})-i \,\arcsin y(r),
  &
    \mbox{if }   r_{1}< r <r_{2}
  \end{array}
$$
where $$y(r)=\mid R_{2}\mid \sin f \frac{-\mid R_{1}\mid^{2}+ \mid R_{1}\mid
\mid R_{2}\mid \cos f+\sqrt{r^{2}R^{2}-\mid R_{1}\mid^{2}\mid R_{2}\mid^{2}\sin
^{2} f}}{rR^{2}},
$$
with $f$ being the phase difference between $R_{1},\overline{R_{2}}$. Similar
expressions exist for $A(r,R_{1},R_{1})$ and  $A(r,R_{2},R_{2})$. With account
of obvious identity
$$
((\lambda^{2}+r_{1}^{2}+R_{1}\overline{R_{2}})^{2}-4r_{1}^2R_{1}\overline{R_{2}})-
$$
$$
((\lambda^{2}+r_{2}^{2}+R_{1}\overline{R_{2}})^{2}-4r_{2}^2R_{1}\overline{R_{2}})=
$$
$$
(R^{2}+2 i \sin{f}\mid R_{1}\mid \mid R_{2} \mid)
\sqrt{\frac{1}{4}-\frac{\lambda^{2}}{R^{2}}} (\mid R_{1}\mid^{2}- \mid R_{2}
\mid^{2}))$$ and integration in (7) we obtain  finally for the overlap integral
\begin {equation}
S_{12}=\exp[-\frac{\pi S}{c^{2}}(R_{12}^{2}-4i\Delta _{12})],
\end {equation}
if $R_{12}< 2\lambda$ or
\begin {equation}
S_{12}=\exp[-\frac{S\pi}{c^{2}}( R_{12}^{2}-4i\Delta _{12}
-R_{12}\sqrt{R_{12}^2- 4 \lambda^{2}}+2
\lambda^{2}ln(\frac{R_{12}+\sqrt{R_{12}^{2}-4
\lambda^{2}}}{R_{12}-\sqrt{R_{12}^{2}-4 \lambda^{2}}})],
\end {equation}
if $R_{12}> 2\lambda $. As above in expression (3) here
$
\Delta _{12}
$
is an area of a sector skyrmion  covers while moving in the plane. Skyrmion,
having wandered closed contour on the plane, acquires the phase $4\pi
S\frac{\Delta _{12}}{c^{2}}$, that is  the skyrmion accumulates a phase of
$4\pi S$ for every spin it encircles.  So, its quantum motion looks  like that
of charged particle with unit charge in a uniform magnetic field of strength $
\frac{4\pi\hbar S}{c^2}$ \cite{Stone}, or particle which experiences a
transverse Magnus force $[\vec{ b} \times \vec{ v}],$ where $ \vec{
b}=\frac{4\pi\hbar S}{c^2} \vec{z} $, $\vec{ v}$- velocity, so that
corresponding length scale is $(\frac{4\pi S}{c^2})^{-\frac{1}{2}}$ which is
similar to magnetic length. This is the same  Magnus force that acts on the
classical skyrmion \cite{Volovik} and which is simply twice larger than  one
acting on the out-of-plane magnetic vortex with topological charge
$\frac{1}{2}$ ("half-skyrmion") \cite{NS}.

One should note the specific dependence of the overlap integral on the
spin-site density ($1/c^2$) and the number of spin deviations ($2S$) similar to
the density dependence in expression (3).

When $ R \rightarrow \infty $ the $R$-dependence of the overlap integral obeys
a power low
$$
S_{12}=(\frac{\lambda^{2}}{R^{2}})^{\frac{S2\pi
\lambda^{2}}{c^{2}}}e^{-\frac{S2\pi \lambda^{2}}{c^{2}}}.
$$
The expressions for the overlap of skyrmions were obtained in continuous
approximation which generally is not correct in the vicinity of those values of
displacements $R_{12}>2\lambda $, at which the overlap of distinct spin states
is zero. Nonetheless, direct numerical calculation for large enough 2D lattices
shows that logarithm of the skyrmion overlap integral is ``almost everywhere''
satisfactorily described by the continuous expression (11) except for discrete
set of values of  $R_{12}$, in which logarithm turns to $-\infty $ .

We considered everywhere the overlap of the skyrmions with equal $\lambda$,
that is with equal global phase and radius. It is easy to see, that skyrmion
states with different $\lambda$ and phases were orthogonal:
$S_{12}^{\lambda\varphi\lambda '\varphi ' }\propto \delta _{\lambda \lambda
'}\delta _{\varphi \varphi '}$, which agrees with the "conservation low" of
the quantity $\lambda$ and phase in the skyrmion \cite{Stern}.

In conclusion, we made use a simplified form for the quantum skyrmion wave
function to obtain the analytical expression for appropriate overlap integral,
 which form confirms a known statement \cite{NS,Volovik,Stone} that skyrmion moves like a particle in uniform magnetic field, or particle which experiences a Magnus force.
We hope the making use of more realistic wave function will keep up principal
features of above derived result.

We would like to thank G. Volovik and J. Schliemann   for helpful comments and
N.Mikushina  for useful collaboration.

\end {document}